\documentclass{llncs}
\usepackage{makeidx}  
\usepackage{verbatim}
\usepackage{symbols}
\usepackage{graphicx}
\usepackage{color}
\usepackage[cmex10]{amsmath}
\usepackage{amsfonts}

\begin{document}
\frontmatter          
\pagestyle{headings}  
\addtocmark{Hamiltonian Mechanics} 
\title{
Local State Space Analysis to Assist Partial Order Reduction
}
\titlerunning{Hamiltonian Mechanics}  
%
\author{Hao Zheng\inst{1} \and Yingying Zhang\inst{1}} %
\authorrunning{Ivar Ekeland et al.} 
%
%
\institute{University of South Florida, Tampa FL 33620, USA,\\
\email{zheng@cse.usf.edu, yingyingz@mail.usf.edu}\\
}
\maketitle              

\begin{abstract}
This paper presents an approach to more efficient partial order reduction for model checking concurrent systems.  This approach utilizes a compositional reachability analysis to generate over-approximate local state transition models for all processes in a concurrent system where an independence relation and other useful information can be extracted.  The extracted independence relation, compared to what can be obtained by statically analyzing the system descriptions, is more precise and refined, therefore leads to more efficient partial order reduction.   This approach is demonstrated on a set of concurrent system examples.  Significantly higher reduction in state space has been observed in several cases compared to what can be obtained using SPIN.
\end{abstract}

\section{Introduction}
\label{intro}

This paper considers an approach to the state explosion problem in model checking concurrent systems.  A concurrent system typically consists of a number of processes running asynchronously.  The communication and synchronization are accomplished by either message passing or shared memory.  When verifying such a system, concurrently enabled process executions need to be interleaved so that all possible orderings of executions are considered to avoid missing any behavior.  The need to consider all possible interleavings of concurrent executions is the main cause of state explosion as the number of interleavings grows exponentially if a system has a high degree of concurrency, and this leads to an excessively large state space for even a relatively small system.

In practice, specifications often are not sensitive to the different orderings of concurrent executions. 
In order to address state explosion due to the interleavings of concurrent executions, various partial order reduction (POR) methods have been developed \cite{Kahlon09,Gueta07,Flanagan05,Clarke99}, which aim to restrict verification to a (significantly smaller) subset of behaviors of a system with guarantee of soundness.  In other words, the ignored behaviors do not carry any new information and do not affect the final verification results.  The general idea of POR is that a subset of all enabled executions in a state is computed for verification, while the ignored ones are deemed irrelevant.  If the computed subset is much smaller that the full set of enabled executions for many states, it can lead to an exponentially smaller state space while still representing sufficient behavior for verification to derive the same results as on the original full state space.
Identifying unnecessary interleavings among the enabled executions plays an important role in POR which is based on examining the dependency relations that exist between the executions of a system.  A dependence relation represents possible interference among concurrent executions, and it is used to identify whether different orderings of concurrent executions are relevant to the specification under verification. 

The efficiency and effectiveness of POR relies on the precision of the derived dependence relations.  Traditionally, they are derived by applying static analysis a priori to the system descriptions by detecting any potential interference between any two executions.  As computing the precise dependency relations between executions may be as hard as verifying the whole system itself, a conservative statically computed approximation is used by POR \cite{Patrice95,Clarke99}. Dynamic POR \cite{Flanagan05,Gueta07}, which excludes the need to apply static analysis a priori by detecting dependencies on-the-fly, and previous work shows that these dynamic approaches achieve higher reduction due to more precise dependences obtained.

This paper presents an alternative approach to computing (in)dependence relations.  It is similar to the static analysis approach in that the dependence relations are computed a priori.   In this approach, a compositional state space construction method \cite{Zheng10} is applied to a concurrent system to find the local state transition model for each consisting process, and subsequently dependence relations are obtained by analyzing these local state transition models.  As the generated local state transition models over-approximate the concrete state space of the processes in the system, the derived dependence relations also hold in the non-reduced state space of the whole system.  This ensures the soundness of this approach.  Once the necessary information is gleaned, model checking with POR can be performed as usual.  The dependence relation found this way is more refined and accurate compared with the traditional static analysis based approaches, and this leads to more significant reduction than allowed by the static approaches.  After a dependence relation is obtained, another key step is to produce a subset of all enabled executions in every state.  This paper follows the correctness conditions of the ample set method \cite{Clarke00} to produce such subsets.  Ideally, the ample set generated in each state should be minimized to achieve the maximal reduction.  To meet this requirement, this paper also show how other useful information can be extracted from the local state space analysis.

The main contribution of this paper is the use of local state space analysis to assist POR to achieve much higher reduction during model checking compared with the static analysis based approaches.  This improvement is due to the more precise and refined dependence relations that can be extracted from the local state transition models.  To the best of our knowledge, this is the first work that combines the compositional state space construction with the dependence relation computation for POR.  The presented approach is implemented in a prototype, and tried on a small set of examples.  The experimental results show definitive improvements over the POR implemented in the SPIN.  

In this paper, it is assumed that the state space models considered have cycles, and model checking with POR used is stateful.  Also, process executions are also referred to as transitions, a term commonly used in POR literatures.

\section{Background}
\label{background}

\subsection{State Graphs}
\label{sg}

A concurrent system is typically described in some high-level language.  Since the exact formalism for describing a concurrent system is not important, this paper simply assumes that a system is described with an initial state $\initst$ and a set of transitions $T$ defining how the system changes its states when transitions in $T$ are executed.   Each transition in $T$ is predicated with a guard defining when it becomes enabled, and one or more actions defining how state variables are changed once the transition is executed.  Furthermore, given a system state $s$, the set of transitions enabled in state $s$ is denoted by $enabled(s)$.  The successor state $s^{\prime}$ after executing a transition $t$ is denoted by $t(s)$.

This paper uses \emph{state graphs} (SGs) to represent the state transition level semantics of these systems.
The definition of state graphs is given as follows.  First, Let $\mathbb{Z}$ be the set of integers.

\begin{Definition}
\label{stg-def}
A state graph $\STG$ is a tuple $(V, Dom, P, \stateset, \initst, \TR, L, F)$ where
\begin{enumerate}
\item $V$ is a finite set of discrete variables.

\item $Dom: V \rightarrow 2^\mathbb{Z}$ is a function that specifies a finite domain for each variable in $V$.

\item $P = \{ (v, i) ~|~v \in V \mbox{ and } i \in Dom(v)\}$ is a finite set of atomic propositions on the variables in $V$.

\item $\StateSet$ is a finite non-empty set of states,  

\item $\initst \in \stateset$ is the initial state.

\item $\TR \subseteq \stateset \times \stateset$ is the set of state transitions.  A state transition is the result of executing a transition in $T$. 

\item $L: \StateSet \rightarrow (V \times \mathbb{Z})^{|V|}$ is a labeling function that associates each state with a set of atomic propositions, one for each variable in $V$. 

\end{enumerate}
\end{Definition} 

\begin{figure*}[tb]
\begin{center}
\begin{tabular}{cc}
\begin{minipage}{3.4in}
\includegraphics[width=3.2in]{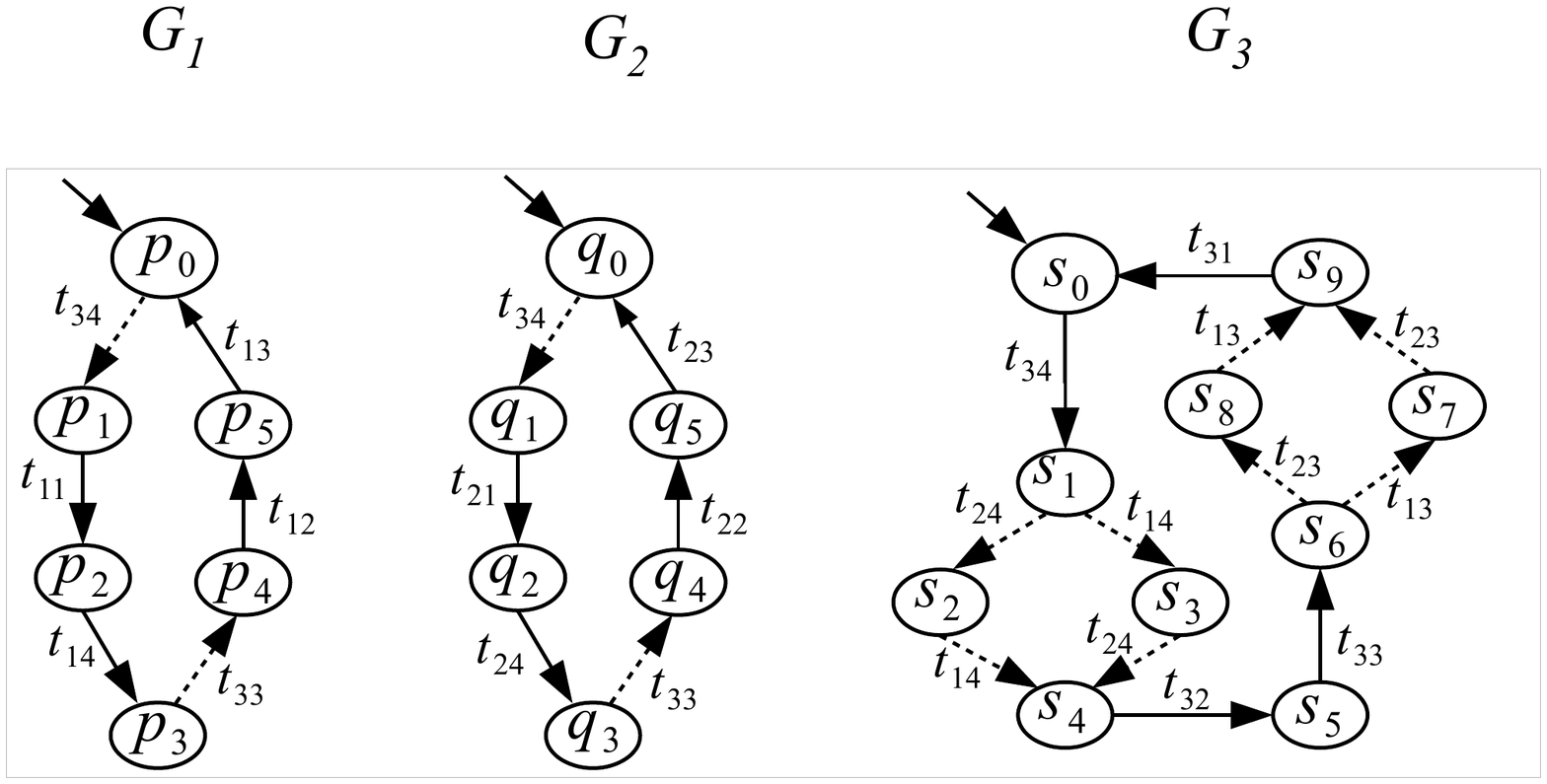}
\end{minipage}
&
\begin{minipage}{1.2in}
Local Variables
\[
\begin{array}{ccc}
G_1 & : & v \\
G_2 & : & w \\
G_3 & : & u
\end{array}
\]
Shared variables
\[x, y, z\]
\end{minipage} \\
\end{tabular}
{\caption{\label{example} An example of a simple concurrent system with three processes.  The local state graphs for the processes are shown in this figure labeled under $G_{1}$, $G_{2}$, and $G_{3}$, respectively.  The local and shared variables in this example are shown on the right hand side of the figure.  Transitions are labeled to indicate how states are changed.  The labelings also make referencing to the transitions easier.  For each $a \in \{u, v, w, x, y, z\}$, $a+$ means that $a$ changes to $1$ from $0$, and $a-$ changes $a$ to $0$ from $1$.}}
\end{center}
\end{figure*}

A large and complex system usually consists of processs connected in a network where communications among the processs can be done through {\em shared variables}.   It is possible to construct process SGs first, and then the SG of the entire system by composing the process SGs asynchronously.  Let $\stg_{i} = (V_i, Dom_i, P_{i}, \StateSet_i, \initst_i, \TR_{i}, L_{i}, F_{i})$, $i \in \{1, 2\}$, be two process SGs, and $C \subseteq P_1 \cap P_2$ be the set of all propositions on the shared variables $V_1 \cap V_2$ between $\stg_1$ and $\stg_2$.  The Asynchronous parallel composition $\stg_1 \| \stg_2$ is a SG $\stg = (V, Dom, P, \stateset, \initst, \TR, L, F)$ such that
\begin{enumerate}
\item $V = V_{1} \cup V_{2}$,

\item $Dom = Dom_1 \cup Dom_2$,

\item $P = P_{1} \cup P_{2}$.

\item $\stateset = \{ (\State_{1}, \State_{2}) \in \stateset_1 \times \stateset_2~|~ L_{1}(\State_{1}) \cap C = L_{2}(\State_{2}) \cap C$ \} 

\item $\initst = (\initst_1, \initst_2)$.

\item $\TR \subseteq \stateset \times \stateset$ such that all the following conditions hold.  For each $((s_1, s_2), (s^\prime_1,s^\prime_2)) \in \TR$,
	\begin{enumerate}
	\item $F_{1}(s_{1}) = 0$ and $F_{2}(\State_{2})=0$.
	\item $L_1(s_1) \cap C = L_1(s^\prime_1) \cap C \imply \TR_1(\state_1, \state^\prime_1) \wedge \state^\prime_2 = \state_2$,     
	\item $L_2(s_2) \cap C = L_2(s^\prime_2) \cap C \imply \TR_2(\state_2, \state^\prime_2) \wedge \state^\prime_1 = \state_1$,
	\item $(L_1(s_1) \cap C \not= L_2(s_2) \cap C) \wedge (L_1(s^\prime_1) \cap C \not= L_2(s^\prime_2) \cap C) \imply \TR_1(\state_1, \state^\prime_1) \wedge \TR_2(\state_2, \state^\prime_2)$,
	\end{enumerate}

\item $L = \{L_{1}(\State_{1}) \cup L_{2}(\State_{2}) ~|~ (\State_{1}, \State_{2}) \in \StateSet\}$.  

\end{enumerate}

In the above definition, when several processs execute concurrently,  they synchronize on the changes of the shared variables, and proceed independently otherwise.  
n the actual implementation, when composing two SGs, a reachability analysis algorithm is performed from the initial composite state following the definition for transition relation $\TR$, and therefore, the resulting composite SG contains only the reachable states.

\begin{figure*}[tb]
\begin{center}
\includegraphics[width=3.5in]{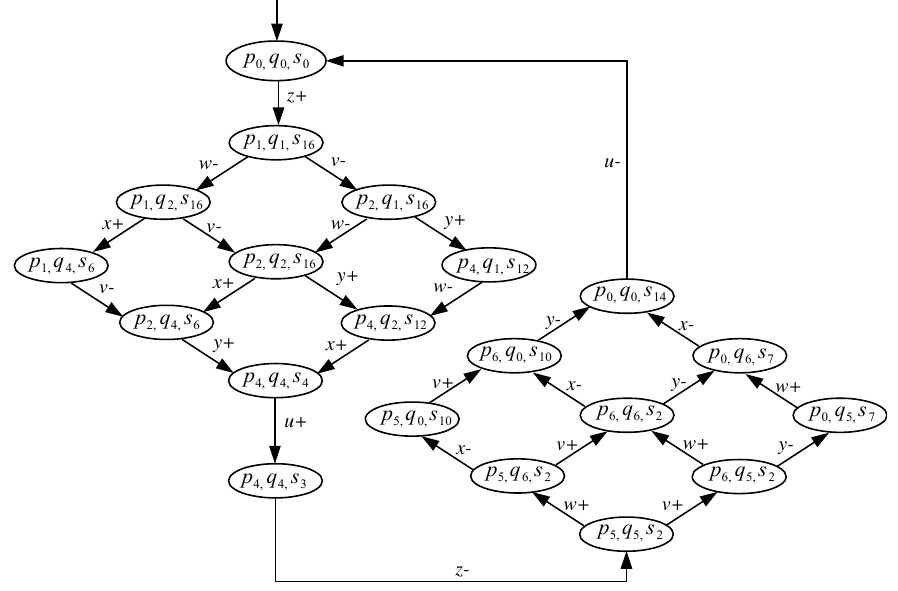}
{\caption{\label{sg-comp} The composite state graph from those in Fig.~\ref{example}.}}
\end{center}
\end{figure*}

Fig.~\ref{example} shows a simple example of a concurrent system with three processes.  The local state graphs for the processes are shown in this figure labeled under $G_{1}$, $G_{2}$, and $G_{3}$, respectively.  The local and shared variables in this example are shown on the right hand side of the figure.  For clarity, the state labelings are not shown.  Instead, state transitions are labeled to indicate how states are changed.  The labelings also make referencing to the transitions easier.  For each $a \in \{u, v, w, x, y, z\}$, $a+$ means that $a$ changes to $1$ from $0$, and $a-$ changes $a$ to $0$ from $1$.  These three SGs can be composed together to form the SG for the whole system as shown in Figure~\ref{sg-comp}.

\subsection{Reachability Analysis}
\label{reachability}

A basic approach for analyzing the dynamic behavior of a concurrent system is reachability analysis, which finds all possible state transitions and thus reachable states for such a system.  The reachable state space is typically represented by a state graph, and can be found by exhaustively executing every enabled transitions starting from the initial state.  A general reachability analysis procedure is given in Algorithm~\ref{basicSearch}.
Checking safety properties can be easily combined with the given algorithm, and liveness properties can be checked using nested-DFS~\cite{Holzmann94}.

\begin{algorithm}[htb]
\caption{\label{basicSearch}Depth first search for reachability analysis}
\KwIn{A system description}
\KwOut{Its state graph}
\BlankLine
$\stateset = \stateset \cup \initst$\;
$stateStack.push(\initst)$\;
$enabledStack.push(enabled(\initst))$\;
\While{stateStack is not empty}{
		$s = stateStack.top()$\;
		$E_s = enabledStack.top()$\;
		\If{$E_s = \emptyset$} {
			$stateStack.pop()$\;
			$enabledStack.pop()$\;
			{\bf continue}\;
		}
		Select $t \in E_s$ to execute\;
		$E_{s = }E_s \backslash t$\;
		$\lSt^\prime = t(\lSt)$\;
		$\TR = \TR \cup \{(s, s^{\prime)}\}$\;
		\If{$\lSt^\prime \notin stateTable$}{
			$stateStack.push(\lSt^\prime)$\;
			$enabledStack.push(enabled(\lSt^\prime))$\;
			$\stateset = \stateset \cup s^\prime$\;		
		}
}
\end{algorithm}

\section{Computing Independence Relations}
\label{Indep}

The major bottleneck for monolithic approaches such as the one shown in Algorithm~\ref{basicSearch} for handling large and complex systems is the excessively large number of states generated due to that all possible orderings of the enabled transitions need to be considered If several are enabled in a state.  On the other hand, it has been discovered that most of the transition interleavings are often irrelevant to the properties under verification.  Therefore, the goal of POR is to identify and remove the redundant interleavings as much as possible.  To make sure that the ignored behaviors do not carry different information for verification, POR relies on transition independence relations.  

In general, the more precise an independence relation is, more redundant interleavings can be identified, thus more efficient POR is.  As indicated at the beginning of this paper, traditional static analysis approaches to computing independence relations have to be conservative for soundness, however they often result in less efficient POR.  This section presents an alternative approach that, instead of based on static analysis of the structures of system models, extract independence relations by analyzing the local state graphs of the consisting processes in a system.  Since the local state graphs capture all runtime behaviors of the processes, the obtained independence relations are much more precise and refined that what can be obtained by static analysis.

This section first reviews the general requirements of independent transitions. It then reviews a compositional SG construction method \cite{Zheng10} for finding local SGs, and shows how an independence relation can be extracted from these local SGs. 

\subsection{General Definition of Independence Relations}
\label{Indep_def}


The state space of a concurrent system often contains many execution paths that correspond to different execution orderings of transitions in different parts of the system \cite{Holzmann94}. If transitions are independent, then their executions do not interfere with each other, which means that changing the execution orderings does not modify their the system behavior \cite{Flanagan05}.  In this case, it is sufficient to consider only one execution ordering.  The notion of independence transitions similar to those in \cite{Holzmann94,Patrice95,Clarke00} is formalized in the following definition.  
	\begin{Definition}
	\label{IndepDef}
 		An independence relation $I \subseteq T \times T$ is a symmetric and antireflexive relation over transitions in $T$ such that for each $(t_1, t_2) \in I$, the following conditions need to hold in every state $s \in S$.
        \begin{enumerate}
        \item If $t_1, t_2 \in enabled(s)$, then $t_2 \in enabled(t_{1}(s))$ and $t_1 \in enabled(t_{2}(s))$,
        \item If $t_1, t_2 \in enabled(s)$, $t_1(t_2(s)) = t_2(t_1(s))$,

        \end{enumerate}
       The dependence relation $D = T \times T - I$. 
	\end{Definition}

Intuitively, independent transitions cannot disable each other, and the executions of independent transitions in different orders are indistinguishable to the specification.  Additionally, the independence is defined for transitions enabled in a state.  It is possible for two transitions not enabled together in every state. In a state where they are not both enabled, they are regarded as independent in that state by the definition.  This definition, however, do not affect the verification results as partial order reduction is only applied to independent transitions that are both enabled.

\subsection{Compositional State Space Construction}
\label{cra}

Given a concurrent system consisting of with a number of processes, this section reviews a compositional local SG construction method for finding SGs for the processes in a system in a high-level description \cite{zheng-tcad10}.  This paper assumes that a system is described as a network of concurrent processes communicating through shared variables.  When generating the local SGs, the interleavings of the invisible state transitions from different processes are not considered, therefore the complexity of generating the local SGs is much lower than that for considering the whole system at the beginning.  

Consider a concurrent system $M = M_1 || M_2 ||...|| M_n$ which is the parallel composition of processes $M_i(1 \leq i \leq n)$ described in some high-level description language such as Promela.  Moreover, every process is assumed to share some variables with some other processes.  The compositional state space construction method builds the SG $\stg_i$ for each process $M_i~(1 \leq i \leq n)$ from an under-approximate context, and gradually expands its SG by including all state and transitions allowed by its neighboring processes.   The main idea is as follows.  Initially, the shared variables that a process depends on are fixed.  Then, this method iteratively performs the following two tasks in each step.
\begin{enumerate}
\item For every process, use the traditional reachability analysis to find all states and state transitions allowed by its definition.
\item For every process, if the above step finds state transitions that cause changes on the shared variables that other processes depend on, extract constraints from these transitions, and apply these constraints to those processes depending on these shared variables.  Applying constraints to a process SG would introduce additional states and state transitions to simulate the synchronization in the SG parallel composition definition in section~\ref{sg}. 
\end{enumerate}
The above two tasks are performed repetitively until no new state or state transitions can be found for any process SG.  

The key to this method is the representation of the changes on the shared variables.  First, let $L_i$ and $L_j$ be the state labeling functions for process SGs $\stg_i$ and $\stg_j$, and $C_{ij} = P_i \cap P_j$ be the set of propositions on the shared variables between $G_i$ and $G_j$.  Suppose that the traditional reachability analysis finds a state transition $(s_i, s^\prime_i)$ for a process SG $\stg_i$, and this transition changes some shared variables that another process SG $\stg_j$ depends on.  This method would extract a constraint $(X, Y)$ which is a pair of sets of propositions on these shared variables such that $X = L_i(s_i) \cap C_{ij}$, $Y = L_i(s^\prime_i) \cap C_{ij}$, and $X \not= Y$.  From an external view, the constraint $(X, Y)$ indicates that whenever these shared variables assume the values satisfying the propositions in $X$, process SG $\stg_i$ {\em may} change the values of these variables to satisfy the propositions in $Y$.  From $\stg_i$, a set of such constraints can be extracted in every step.

When applying these constraints to $\stg_j$ that depend on the same shared variables, for each constraint $(X, Y)$, this method checks if there is any state $s_j$ in $\stg_j$ such that $X \subseteq L_j(s_j)$.  If $s_j$ exists, then a new state $s^\prime_j$ and a new state transition $(s_j, s^\prime_j)$ due to this constraint are generated for $\stg_j$ such that $L_j(s^\prime_j) = (L_j(s_j) - X) \cup Y$.  State $s_j^\prime$ and transition $(s_j, s^\prime_j)$ are added to reflect the effect of $\stg_j$'s environment on its behavior.  From these new states, more new states and transitions may be found according to the definition of process $M_j$.  Subsequently, the traditional reachability is applied again. 

\begin{figure}[tb]
\begin{center}
\includegraphics[width=80mm]{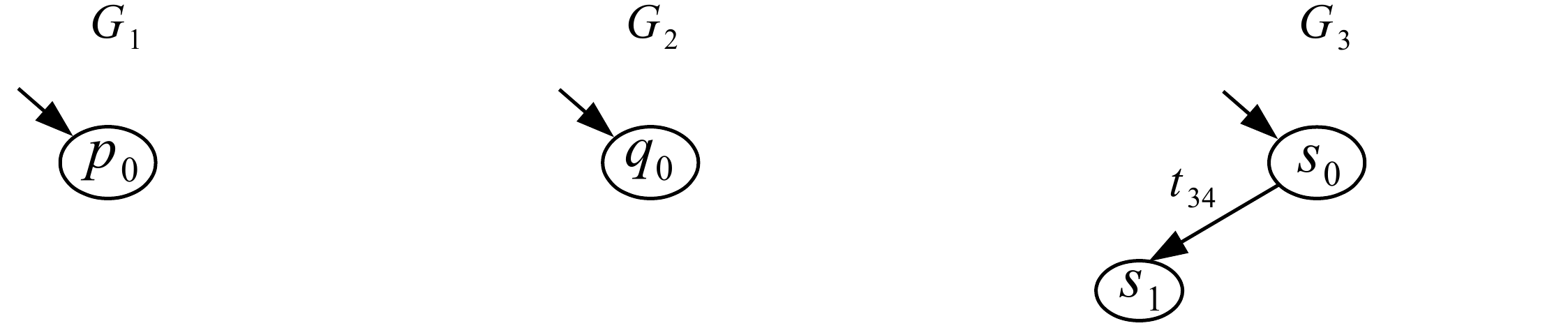} \\ (a) \\ \vspace{2mm}
\includegraphics[width=85mm]{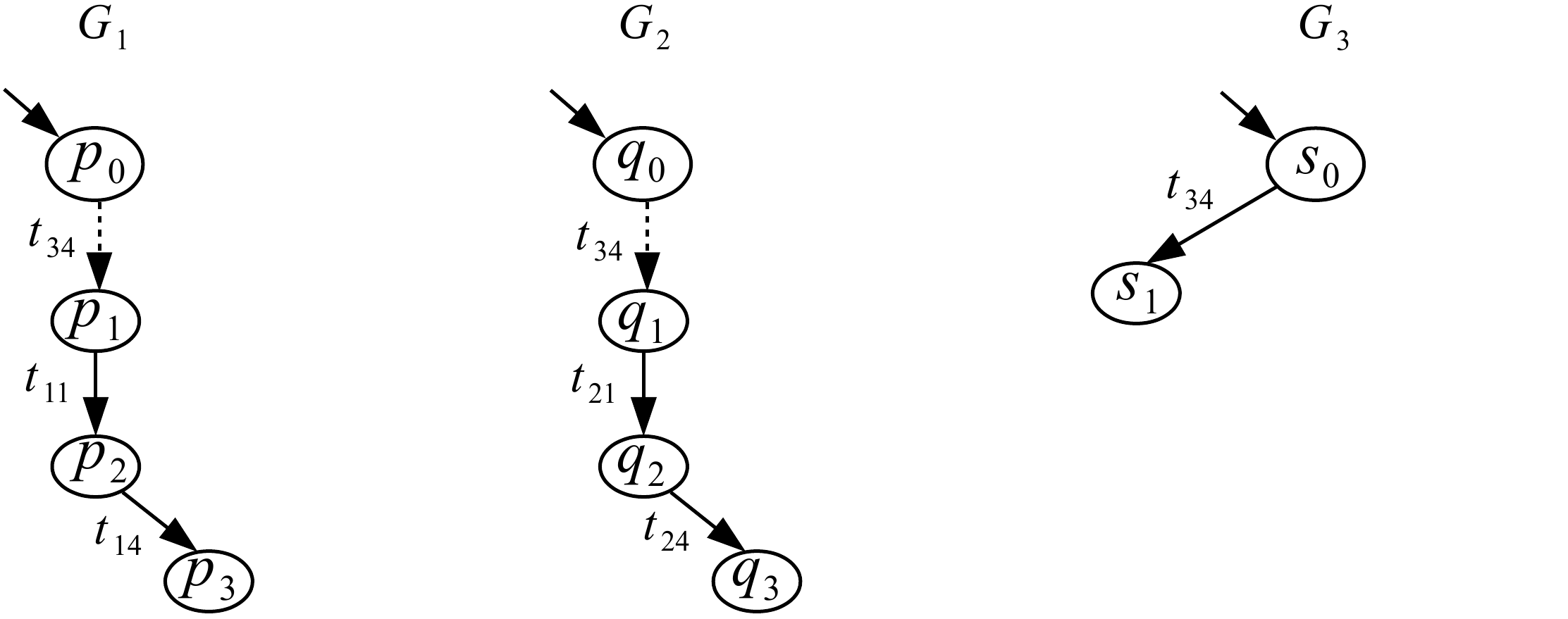} \\ (b) \\ \vspace{2mm}
\includegraphics[width=85mm]{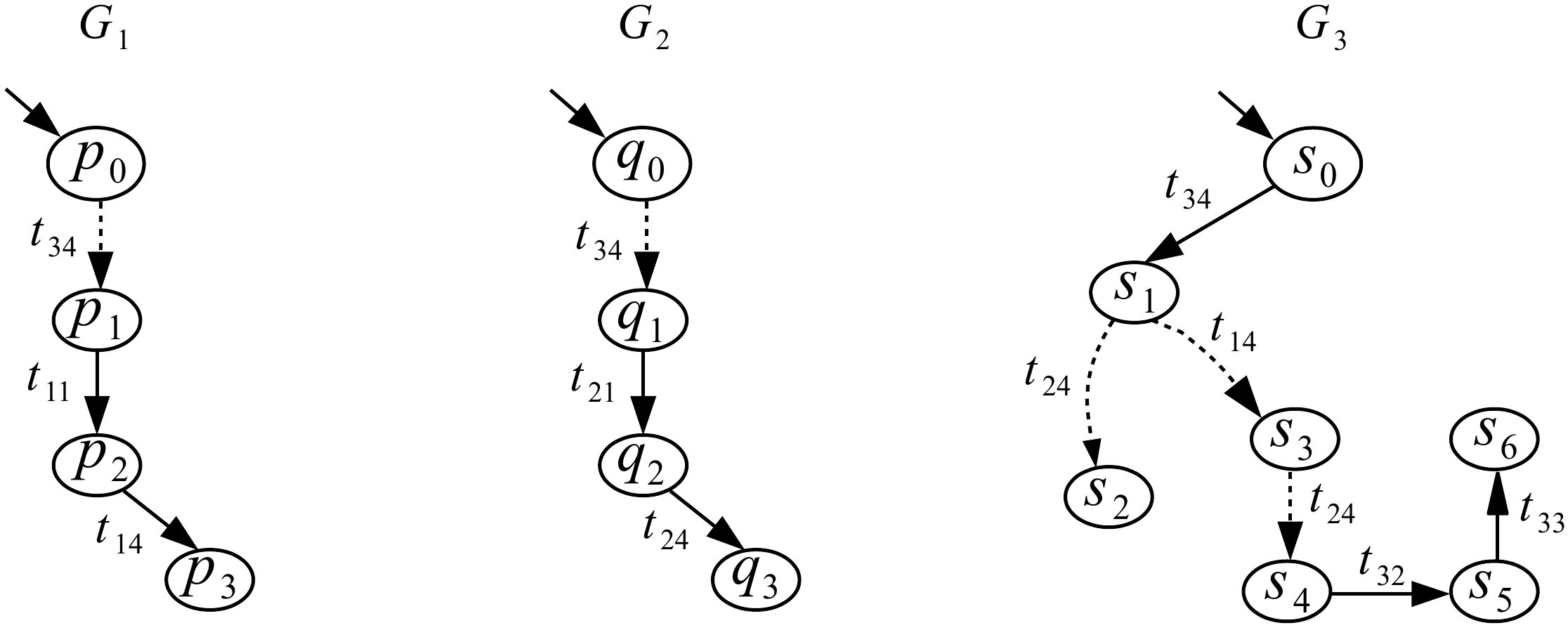} \\ (c) \\ \vspace{2mm}
{\caption{\label{cra-ex} Figures in (a)-(c) show the partial process state graphs generated during the compositional reachability analysis.The final process state graphs generated when the compositional reachability analysis is shown in Fig~\ref{example}.}}
\end{center}
\end{figure}

Now consider the simple asynchronous circuit example as shown in Fig~\ref{ex-circuit}(a) with three processes $M_1$, $M_2$ and $M_3$.  Both $M_1$ and $M_2$ depend on variable $z$ while $M_3$ depends on variables $x$ from $M_2$ and $y$ from $M_1$.   Suppose that the initial states for these processes are defined below, respectively,
\[
\begin{array}{rcl}
L_1(p_0) & = & \{z=0, v=1, x=0\}, \\
L_2(q_0) & = & \{z=0, w=1, y=0\}, \\
L_3(s_0) & = & \{x=0, y=0, u=0, z=0\}. 
\end{array}
\]  
In the first step, no states and transitions can be added to $\stg_1$ and $\stg_2$ as no variables can be changed according to the circuit definition.  For $M_3$, variable $z$ is changed to $1$ from $0$, therefore a new state $s_{16}$ and a new transition $(s_0, s_{16})$ are added to $\stg_3$.   The result is shown in Fig~\ref{cra-ex}(a).  Since the transition $(s_0, s_{16})$ changes $z$ that $M_1$ and $M_2$ depend on, a constraint $(\{z=0\}, \{z=1\})$ is extracted from $\stg_3$, and applied to $\stg_1$ and $\stg_2$ where states $p_1$ and $q_1$ are generated.  After the traditional reachability analysis is applied to all three SGs, more states and transitions are added to $\stg_1$ and $\stg_2$, respectively, while $\stg_3$ remains unchanged as shown in Fig~\ref{cra-ex}(b).  Now, $\stg_1$ has a transition $(p_2, p_4)$ that changes variable $y$ to $1$ from $0$, and $\stg_2$ has a transition $(q_2, q_4)$ that changes variable $x$ to $1$ from $0$.  Since $\stg_3$ depends on both variables, two constraints are extracted from $\stg_1$ and $\stg_2$, respectively,
\[
(\{y=0\}, \{y=1\}),~~~(\{x=0\}, \{x=1\}).
\]
After applying these constraints to $\stg_3$, three new states, $s_6$, $s_{12}$, and $s_4$ are added.  From these new states, the traditional reachability analysis finds two new states $s_3$ and $s_2$ as defined in the SG $G_{3}$ in Fig.~\ref{example}.  The results after this step is shown in Fig~\ref{cra-ex}(c).  Eventually, all states and transitions are found for these three processes, and their SGs are shown in Fig~\ref{example}.

In the end, the generated process SGs are an over-approximations of the concrete process SGs if they were produced when the whole design was considered.  This means that the process SGs may contain states and state transitions that do not exist in the global SGs obtained by applying reachability analysis directly to the whole system description.  This is proved in \cite{zheng-tcad10}.  This property of the generated process SGs guarantees the soundness of the approach presented in this paper.

\subsection{Extracting Dependence Relations}
\label{analysis_approach}

According to Definition~\ref{IndepDef}, the dependence relation is the complement of the independence relation. Therefore,  after the process state graphs are generated, transition dependence relations are found by checking if there is a violation of any condition in Definition~\ref{IndepDef} in any state in any process state graph.  More specifically, two transitions $t_{1}$ and $t_{2}$ are dependent with each other if in any process state graph, there exists a state $\lSt$ such that one of the following conditions holds. 
  	\begin{enumerate}
  	\item $\exists (s, t_1, s^\prime):  t_2 \in enabled(s) \wedge t_2 \notin enabled(s^\prime)$,
  	\item $\exists (s, t_2, s^\prime):  t_1 \in enabled(s) \wedge t_1 \notin enabled(s^\prime)$ ,
  	\item $t_1, t_2 \in enabled(\lSt) \wedge t_1(t_2(s)) \not= t_2(t_1(s))$,
	\end{enumerate}

\begin{figure}[tb]
\begin{center}
\includegraphics[height=4cm]{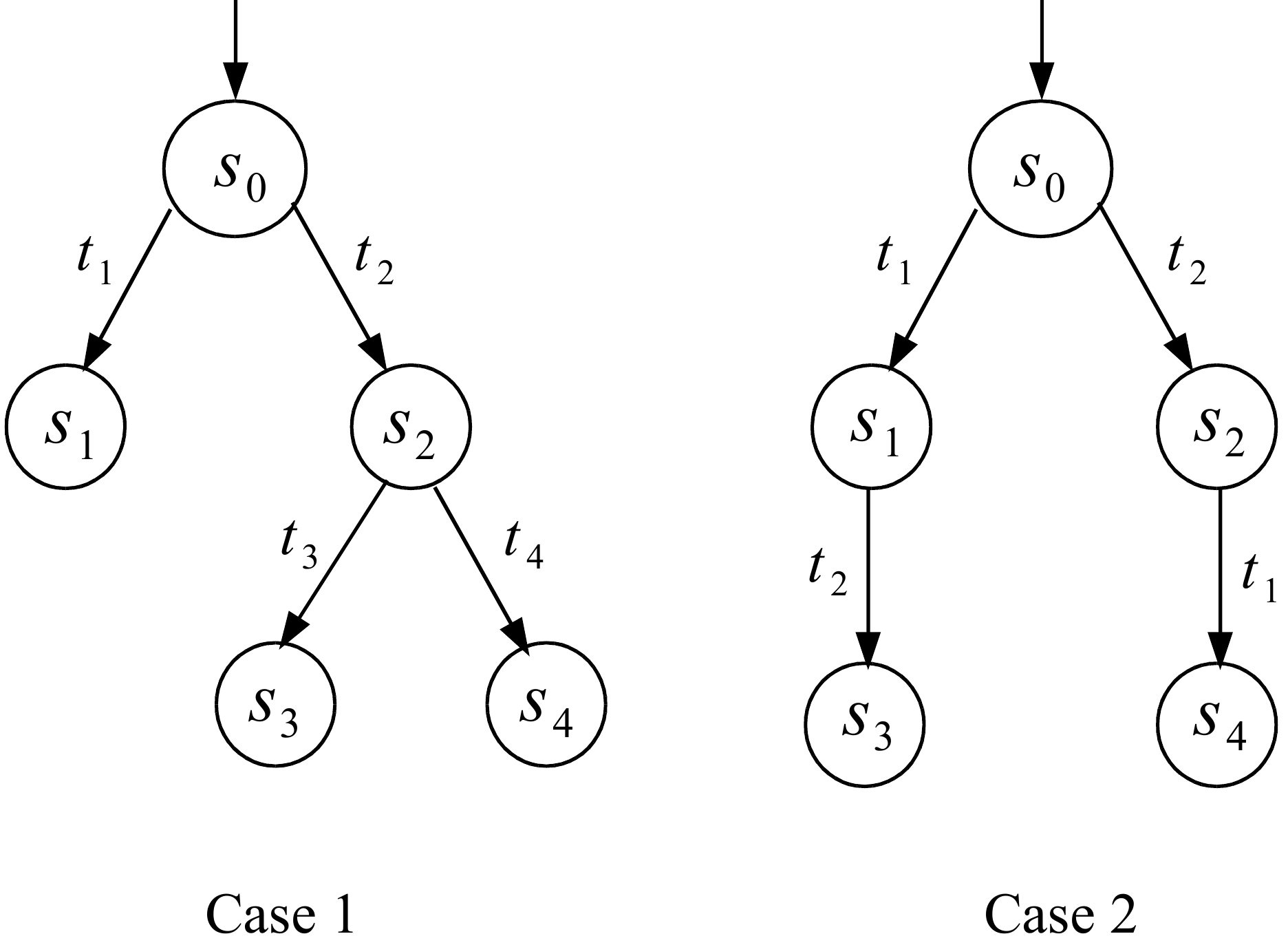}
\caption{\label{dep_CRA} Examples of dependent transitions that can be found in the state graphs as defined in Definition~\ref{IndepDef}.}
\end{center}
\end{figure}
In the above definition, two transitions depend on each other if firing one can disable the other one as stated in the first two conditions, or firing these two transitions in different orders leads to different states.  Fig~\ref{dep_CRA} illustrates two cases of dependent transitions found when one of the above conditions holds.  In Case~1 shown in~Fig~\ref{dep_CRA}, $t_{1}$ and $t_{2}$ are dependent because condition~1 or 2 is violated, while in Case~2, condition~3 is violated.  Note that in this method if any of these conditions holds for transitions in any state in any process state graph, they are regarded as dependent globally.  

In this local SG analysis based approach, much more accurate information on how different orderings of firing transitions affect, for example, whether a transition can possibly be disabled by another transition, is readily available, therefore much more refined  dependence relations, and equivalently independence relations, can be derived.  This allows many transitions enabled in each state to be removed during state space search, thus leading to enormous reduction in state space, as shown by the experimental results in section~\ref{results}.  On other hand, the local state graphs generated are over-approximations as these models may contain additional states and state transitions that would not exist if the monolithic state graph is generated for the whole system as shown  in \cite{zheng-tcad10}.  Because of these additional states and state transitions, dependence relation found may still be conservative.  This is equivalent to saying that some transitions that should be independent may be found as being dependent due to the additional states and state transitions.  However, this does not affect the correctness of this method, and it would only cause partial order reduction to be less effective.

\section{Partial Order Reduction}
\label{POR}

First of all, it is necessary to differentiate the global states of the whole design and the local states of the processs of the design.  Typically, a concurrent system consists of processes, each of which makes changes to variables responding the input changes caused by its neighboring processes. Let $M = M_1 \| \ldots \| M_n$ be the description of a system where $M_i (1 \leq i \leq n)$ is the description for the $i$th process, and $\|$ is the parallel composition operator appropriately defined for such a description.  The states of the whole system $M$ are referred to as the {\em global} states denoted $\gSt$, and the states of the individual modules are referred to as {\em local} states denoted $\lSt$.  The global states for $M$ are not represented as monolithic entities either.  Instead, they have similar structure to that of $M$.  More specifically, a global state $\gSt$ of $M$ is a $n$-tuple of the local states $(s_1, \ldots, s_n)$ where each $s_i~(1 \leq i \leq n)$ is a local state of LPN module $M_i$.  The shared variables are replicated in the state of the local SGs where they belong.  The global states are found when the state space of the whole system is searched, while the local states are generated during the compositional state space construction as described in the last section.  For convenience, a predicate $in(\gSt, \lSt_i)$ is defined to indicate whether the local state $\lSt_i$ is part of the global state $\gSt$.  $in(\gSt, \lSt_i)$ holds  if the local state $\lSt_i$ is part of the global state $\gSt$.

After the independence relation is obtained, the next step is to compute a subset of the enabled transitions in each state for POR during the state space exploration.   This paper considers ample set method as described in \cite{Clarke00}.  Let $ample(\gSt)$ denote the ample set computed for a global state $\gSt$.  To reduce the complexity of the state space exploration, the size of $ample(\gSt)$ should be as small as possible, and computing $ample(\gSt)$ should have low overhead.  To preserve the sufficient behavior to verify all properties soundly, four conditions need to be satisfied for ample set construction as described in~\cite{Clarke00}, which are listed below for convenience.

\begin{description}
\item[C0] $ample(\gSt) = \emptyset$ iff $enabled(\gSt) = \emptyset$.
\item[C1] In the full state graph, on any path starting from a state $\gSt$, a transition dependent on a transition in $ample(\gSt)$ cannot occur before some transition from $ample(\gSt)$ occurs first.
\item[C2] If a state is not fully expanded, every transition in the ample set is inisible.
\item[C3] If a cycle contains a state in which some transition $t$ is enabled, then it also contains a state $\gSt$ such that $t \in ample(\gSt)$.
\end{description}

Among all four conditions, {\bf C0} and {\bf C2} can be readily satisfied.  Some more discussion on {\bf C3} is givne in a later section.  This section considers {\bf C1} for computing $ample(\gSt)$ as it is the most difficult one.



\subsection{Computing Ample Set}
\label{indep_pro}

\begin{figure}[tb]
\begin{center}
\includegraphics[width=60mm]{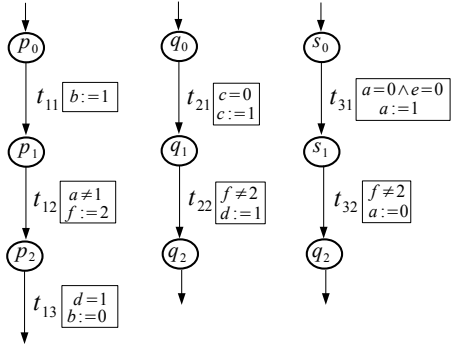}
\caption{\label{ampleP}An example showing that just considering local independence for computing ample set is not enough.}
\end{center}
\end{figure}

This section presents two conditions to make sure that all essential interleavings between dependent transitions are preserved.  

\noindent{\bf Condition 1} For all enabled transitions $t_1, t_2 \in enabled(\gSt)$, if $t_1 \in ample(\gSt)$ and $t_2 \not\in ample(\gSt)$,  then $I(t_1, t_2)$ holds.

Intuitively, Condition~1 requires that every transition in $enabled(\gSt) \backslash ample(\gSt)$ is independent on every transition in $ample(\gSt)$. This condition guarantees that transitions in $enabled(\gSt) \backslash ample(\gSt)$ are still enabled after any transition in $ample(\gSt)$ is fired.  However, this condition alone is not sufficient to guarantee that no states can be missed by firing transitions in $ample(\gSt)$ first. Refer to Fig.~\ref{ampleP} for a simple example. In the initial state $\gSt_0 = (p_0, q_0, s_0)$, $enabled(\gSt_0) = \{t_1,t_2,t_4\}$. Suppose that every transition in $enabled(\gSt_0)$ is independent with each other. Therefore, any single transition can be chosen to be included in $ample(\gSt_0)$ according to {Condition~1}, and it can be seen that the other transitions can still remain enabled after firing the transition in $ample(\gSt_0)$.  First, choose $t_1$ to be included in $ample(\gSt_0)$.  Suppose that another state $\gSt_1 = (p_k, q_0, s_0)$ is reached after firing $t_1$ and some other transitions where $t_k$ and $t_4$ are enabled.  From Fig.~\ref{ampleP} it can be seen that firing $t_k$ and $t_4$ in different orders may lead to different state spaces since firing $t_4$ first disables $t_k$.  Now suppose $ample(\gSt_0)=\{t_4\}$ and firing $t_1$ is delayed.  This is still valid according to Condition~1.  Since transition $t_4$ fires before $t_1$,  $t_k$ may never get a chance to fire, thus possibly causing certain states not to be found.  To avoid this situation, the following condition needs to be satisfied as well when computing the ample set.

\noindent{\bf Condition 2}~For each transition $t \in enabled(\gSt)$,  if there is another transition $t^\prime \not\in enabled(\gSt)$ such that $D(t, t^\prime)$ holds, then $t \not\in ample(\gSt)$.

Intuitively, this condition requires that firing an enabled transition, if it is dependent with any other transition that may be enabled in the future, needs to be delayed until its dependent transition also becomes enabled in the same state.  This makes sure that the interleavings involving a currently enabled transition and another one to be enabled in the future are not lost. In the example shown in Fig.~\ref{ampleP}, since $D(t_4, t_k)$ holds for $t_k \notin enabled(\gSt_0)$, in the initial state $\gSt_0$, $t_4 \in ample(\gSt_0) $.  In other words, with {Condition~2}, firing $t_4$ needs to be delayed until both $t_4$ and $t_k$ are enabled, therefore firings of $t_2$ and $t_k$ can be interleaved.

The basic idea behind {Condition~2} is that if a currently enabled transition is dependent with another transition that might be enabled in the future, the computed ample set must preserve the interleavings of firing these two transitions whenever possible by putting off firing the currently enabled transition as late as possible.  It is possible that the only ample set that satisfies both conditions is empty. In this case, all the enabled transitions have to be included in the ample set, i.e. if $ample(\gSt)=\emptyset$ according to Condition~1 and 2, then $ample(\gSt)=enabled(\gSt)$.



An ample set that meet the above two conditions includes only the transitions that are globally independent with any other transitions, thus satisfying the condition described in {\bf C1} trivially.  This implies an ample set satisfies {\bf C1} if it satisfies both condition 1 and 2 defined in this section.

Based on the above discussion, $ample(\gSt)$ can be computed from $enabled(\gSt)$ as follows.
\begin{enumerate}
\item First, $enabled(\gSt)$ is partitioned to two subsets $dep(\gSt)$ and $indep(\gSt)$.  $dep(\gSt)$ includes all transitions $t$ enabled in $\gSt$ such that $D(t, t^\prime)$ for some transition $t^\prime$, and $indep(\gSt) = enabled(\gSt) \backslash dep(\gSt)$ is the set of transitions enabled in $\gSt$ such that each transition $t \in indep(\gSt)$ does not depend with any other transitions.  If $indep(\gSt) \not= \emptyset$, we can choose any single transition of $indep(\gSt)$ to be included in $ample(\gSt)$.

\item If $indep(\gSt) = \emptyset$, perform this step.  First note that condition~2 only requires that a currently enabled transition cannot be included in the ample set if it is dependent with a transition not currently enabled.  However, it is fine to include both $t_1$ and $t_2$ into the ample set if $t_1, t_2 \in enabled(\gSt)$, and $t_1$ and $t_2$ are only dependent with each other.  Therefore, $ample(\gSt)$ can be found such that every transition in $ample(\gSt)$ is independent with any other transition not in $ample(\gSt)$ including enabled and disabled ones. 

\item If $indep(\gSt)$ is still $\emptyset$, $ample(\gSt) = enabled(\gSt)$.
\end{enumerate}

The above partitioning approach is simple to implement, and its overhead for computing the ample set during the runtime is negligible.  On the other hand, the above two conditions could be overly conservative, and thus might cause the reduction less effective.   Consider the example shown Fig.~\ref{ampleP} again.  Based on the above two conditions, firing transition $t_4$ in state $\gSt_0$ needs to delayed to avoid losing the interleavings between $t_k$ and $t_4$ if $t_k$ could be enabled in the future.  Now suppose there exists a transition $t_i$ between $t_1$ and $t_k$ such that $t_i$ can be enabled only after $t_4$ is fired, and it needs to be fired first before $t_k$ can be enabled.  In this case, transitions $t_k$ and $t_4$ can never be enabled together in any state reachable from $\gSt_0$, and the dependency between $t_k$ and $t_4$ does not actually hold in $\gSt_0$.  Therefore, knowing this information can help to expose more independent transitions for reduction not restricted by Condition~2.  This is commonly known as conditional dependency \cite{Patrice95}.

According to the above discussion, in general, given a state $\gSt$ such that $t_1, t_2 \in enabled(\gSt)$ and $D(t_2, t_3)$ holds for some transition $t_3$, it is necessary to know if another state $\gSt_n$ where $D(t_2, t_3)$ holds is reachable from $\gSt$ in order to generate a smaller ample set in state $\gSt$.  This information can also be computed from the local state graphs generated using the method as described in the last section as follows.
\begin{enumerate}
\item Consider every local state $\lSt_{n}$ where a dependency $D(t_2, t_3)$ holds.  Then, for each predecessor state $\lSt_{n-1}$ of $s_n$,  if either $t_2 \in enabled(\lSt_{n-1})$ or $t_3 \in enabled(\lSt_{n-1})$, add $(s_{n-1}, t_2)$ or $(\lSt_{n-1}, t_3)$ into a set $D_C$. If $enabled(\lSt_{n-1})$ does not include either $t_2$ or $t_3$, it is not considered further.  This is because it is only necessary to consider if states with $t_2$ or $t_3$ enabled can reach a state where the dependency involving either $t_2$ or $t_3$ actually holds, and this dependency is irrelevant to other states, therefore they do not need to be considered.

\item The above step repeats on state $s_{n-1}$ if $s_{n-1}$ exists.
\end{enumerate}
The above procedure is applied to every local state graph for every pair of dependent transitions.  Note that the above procedure traverses paths in local SGs to determine reachable dependencies for a local state.  Since these local SGs are over-approximations of the concrete ones, these reachable dependencies may or may not hold in the global state space of the whole system.  On the other hand, it still guarantees the correctness as no dependencies are ignored.

With $D_C$ computed as described above, Condition~2 can be changed to the one shown below. Since the dependency involving a transition enabled in a state may or may not hold, more enabled transitions can be considered for the ample set, and this could lead to a smaller ample set, therefore more reduction in state space complexity.

\noindent{\bf Condition~$2^\prime$}~~For each transition $t \in enabled(\gSt)$,  if $(s, t) \in D_C$ and $in(\gSt, s)$ hold in a global state $\gSt$, then $t \not\in ample(\gSt)$.

\section{Experimental Results}
\label{results}

\begin{table*}
\begin{center}
\caption{\label{table-1}Comparison between the POR method based on the local state graph analysis and the SPIN POR method (Time is in seconds, and memory is in MBs.). $|\stateset|$ is the numbers of states found at the end of reachability analysis.  Entries filled with $-$ indicates time-out.
}
\begin{tabular}{|c|c||c|c|c||c|c|c||c|c|c|}
\hline
 \multicolumn{2}{|c||}{Designs} & \multicolumn{3}{c||}{Monolithic} & \multicolumn{3}{c||}{POR-SPIN} & \multicolumn{3}{c|}{POR-LSSA}  \\
  \hline\hline
Name & Size & Time & Mem & $|\StateSet|$ & Time & Mem & $|\StateSet|$ & Time & Mem & $|\StateSet|$ \\
\hline\hline
	& 3 & $0.315$ & $2.4$  & $3756$ & $0.015$ & $2.781$ & $3756$ & $0.109$ & $1.116$ & $356$ \\ \cline{2-11}
	& 5 & $8.105$ & $61.538$  & $227472$ & $1.65$ & $71.695$ & $227472$ & $0.281$ & $2.152$ & $5099$ \\ \cline{2-11}
arb	& 7 & $-$ & $-$  & $-$ & $-$ & $-$ & $-$ & $1.825$ & $30.477$ & $90754$ \\ \cline{2-11}
	& 9 & $-$ & $-$  & $-$ & $-$ & $-$ & $-$ & $8.376$ & $99.617$ & $398579$ \\ \cline{2-11}
	& 11 & $-$ & $-$  & $-$ & $-$ & $-$ & $-$ & $122.347$ & $1360.516$ & $4862988$ \\ \cline{2-11}
\hline
\hline
	& 3 & $0.119$ & $4.8$  & $644$ & $0$ & $2.195$ & $644$ & $0.047$ & $3.599$ & $51$ \\ \cline{2-11}
	& 5 & $0.733$ & $16.253$  & $20276$ & $0.08$ & $6.593$ & $20276$ & $0.047$ & $3.999$ & $121$ \\ \cline{2-11}
	& 8 & $199.353$ & $845$  & $3572036$ & $30.2$ & $1087.211$ & $3572036$ & $0.062$ & $1.368$ & $286$ \\ \cline{2-11}
	& 10 & $-$ & $-$  & $-$ & $-$ & $-$ & $-$ & $0.078$ & $2.477$ & $436$ \\ \cline{2-11}
fifoN	& 20 & $-$ & $-$  & $-$ & $-$ & $-$ & $-$ & $0.253$ & $6.579$ & $1666$ \\ \cline{2-11}
	& 50 & $-$ & $-$  & $-$ & $-$ & $-$ & $-$ & $0.086$ & $14.279$ & $10156$ \\ \cline{2-11}
	& 100 & $-$ & $-$  & $-$ & $-$ & $-$ & $-$ & $5.086$ & $81.922$ & $40306$ \\ \cline{2-11}
	& 200 & $-$ & $-$  & $-$ & $-$ & $-$ & $-$ & $32.665$ & $328.889$ & $160606$ \\ \cline{2-11}
	& 300 & $-$ & $-$  & $-$ & $-$ & $-$ & $-$ & $137.048$ & $978.714$ & $360906$ \\ 
\hline
\hline
	& 3 & $3.589$ & $26.1$  & $267,999$ & $0.265$ & $19.706$ & $117270$ & $0.202$ & $1.745$ & $912$ \\ \cline{2-11}
	& 4 & $1235$ & $1032$  & $15.7M$ & $15.5$ & $553.421$ & $4678742$ & $0.25$ & $3.685$ & $4495$ \\ \cline{2-11}
dme	& 5 & $-$ & $-$  & $-$ & $-$ & $-$ & $-$ & $0.437$ & $5.252$ & $15452$ \\ \cline{2-11}
	& 8 & $-$ & $-$  & $-$ & $-$ & $-$ & $-$ & $13.118$ & $174.070$ & $687475$ \\ \cline{2-11}
	& 9 & $-$ & $-$  & $-$ & $-$ & $-$ & $-$ & $49.858$ & $532.353$ & $2471839$ \\ \cline{2-11}
\hline

\hline
tagunit & 48  & $-$ & $-$ & $-$ & $4.37$ & $144.984$ & $786672$ & $0.187$ & $4.897$ & $1103$ \\

\hline
pipectrl & 50 & $-$ & $-$  & $-$ & $-$ & $-$ & $-$ & $0.468$ & $5.670$ & $4610$ \\
\hline
mmu & 55 & $-$ & $-$  & $-$ & $-$  & $-$ & $-$ & $13$ & $312$ & $1863$ \\
\hline
\end{tabular}
\end{center}
\end{table*}

The partial order reduction method described in this paper is implemented in a concurrent system verification tool $Platu$, an explicit model checker implemented in Java.  Experiments have been been performed on a number of examples.  These examples include asynchronous circuit designs from previously published papers \cite{Martin:FIFO,dill:PHD,stevens99relative,Yoneda96usingpartial,Myers:PhD}.  All these examples are experimented with the POR method described in this paper and the one implemented in SPIN~\cite{Holzmann1997}.   This provides a comparison of the approach presented in this paper to the state-of-the-art.  In the experiments, only deadlock-freedom and safety properties are considered.  The state spaces of all examples contain cycles, a form of weak proviso algorithm \cite{Holzmann94} is implemented to fulfill the cycle condition {\bf C4} as shown in section~\ref{POR}.

All experiments are performed on a Linux workstation with a Intel Pentium Dual-Core CPU and 4 GB memory, and the results are shown in Table~\ref{table-1}.  For each example, its state space is searched by monolithic reachability analysis (Monolithic), SPIN with partial order reduction (POR-SPIN), and the approach described in this paper (POR-LSSA).  In the table, the first two columns show each examples and their sizes in terms of the numbers of concurrent processes included in each system description. Since these examples are circuit designs, the variable type is Boolean.  For each method used to find the example's state space, the total runtime, memory used, and the number of states found are shown in columns Time, Mem, and $\stateset$.  Runtime is in seconds, and memory is in MB.  For all examples, a limit of $5$ minutes is imposed. Entries in the table with $-$ indicate the search for that example runs out of time or 2 GB memory space is exhausted.  

From table~\ref{table-1}, it can be seen that SPIN is not able to find any reduction for a number of examples including fifo and arb designs with different numbers of processes.  For other examples, some reduction in state space is found by SPIN, but the reduction is not good enough to allow larger examples to be handled.  On the other hand, this approach is far more efficient as it can finish more and much larger examples.  These results show that the behavioral analysis approach indeed is capable of deriving more accurate independence relations from state space models, thus allowing more transitions to be identified as independent.

The overhead of the compositional reachability analysis is relatively small in total runtime.  Except for mmu, generally $10$ percent of the total runtime is spent on constructing the local state graphs, and deriving the dependence information from these state space models.  For mmu, the overhead of the behavioral analysis is about $35$ percent of the total runtime.  This is due to that many components in this design have quite complex interfaces, and a large number of extra states and state transitions are generated for each component during the compositional readability analysis.

\section{Previous Work}

Persistent set \cite{Patrice95} or stubborn set \cite{Valmari1991} are some of the early work on partial order reduction.  These methods, relying on the notion of transition dependence, compute a sufficient subset of enabled transition in each state visited during state space exploration such that every execution path not in the reduced state space is represented by an equivalent one with respect to the properties to be checked.  An alternative technique, sleep sets \cite{Patrice95}, avoids traversing certain state transitions based on the information on the transition dependence in the current state and the states visited in the past.  The sleep set technique sometimes can reduce certain states, but the persistent method outperforms in reduction of states.  Both can be used together to achieve better performance.   

As pointed out earlier, traditional POR methods rely on the dependence relations obtained by the static analysis on the system descriptions, and the conservative approximation of dependence relations is often produced for the soundness.  The consequence is often larger persistent sets calculated for the reachable states encountered on-the-fly, thus large state space to be explored.  To address this problem, dynamic POR as in \cite{Flanagan05} detects dependency   among transitions on-the-fly, and this dynamically obtained dependence relation can be very precise and often leads to smaller persistent sets.  Dynamic POR \cite{Flanagan05} is expensive because it is a stateless model checking procedure, which traverses a sufficient number of execution paths in a concurrent system.  This is in contrast to the stateful approach implemented in SPIN which aims to traverse sufficient set of states with respect to a given property.  The dynamic POR may be inefficient because the number of execution paths that need to be traversed can be exponential compared to the set of state that need to be traversed.  Plus, the state space of the systems under verification is assumed to be acyclic in \cite{Flanagan05}.  There are also efforts for making the original dynamic POR algorithm stateful \cite{Kahlon09,Yu-spin2008}, and extend it to systems that allow cycles in their state space.

POR has been combined with symbolic methods \cite{Chao-tacas2008,Kahlon09}, but these approaches are for bounded model checking using SAT solvers.

The cycle condition as formulated for the ample set method can be fulfilled by the so called proviso ~\cite{Patrice95} algorithm by fully expanding at least one state in every cycle found during POR state space search.  This proviso is strong as it allows full $LTL_{-X}$ formulas to be checked by POR approaches.  For safety properties, a weak form of proviso, where the full expansion is not required if at least one enabled transition in the ample set does not close a cycle, can be used.  A more sophisticated approach to meet the cycle for the safety properties is based on detecting the maximal terminal strongly connected components~\cite{Patrice95}.  However, its implementation could be more involved than the weak proviso approach.  In \cite{Nalumasu97anew}, a two-phase POR algorithm is described and better reductions are shown on a number of examples compared to the POR combined the proviso.  However, its effectiveness relies on if a system model displays sufficient amount of determinism among concurrent processes; otherwise, its performance could be as bad as that of monolithic search. 

\section{Conclusion}
This paper introduces a new approach to computing independence relations for partial order reduction.
Since the analysis is applied on the state space models, the derived independence relations are more refined and accurate, which lead to more effective partial order reduction as shown by the experimental results. In the future, we plan to integrate this approach with compositional verification in order to scale model checking for even larger designs, and also extend the idea presented in this paper to real-time system verification.  Additionally, this approach relies on an efficient local state graph construction method.  If the constructed local state graphs include too many extra states or state transitions, it may have a negative impact on the precision of the derived dependence relations and the efficiency of the overall POR procedure.  The performance of the local state graph construction procedure itself can also be greatly degraded if the generated local state graphs are too large.  Therefore, a better local state graph construction method is needed in terms of local state graphs generated with least number of extra states and transitions and short runtime.

\section*{Acknowledgment}
This material is based upon work supported by the National Science Foundation under Grant No. 0930510 and 0930225. Any opinions, findings, and conclusions or recommendations expressed in this material are those of the author(s) and do not necessarily reflect the views of the National Science Foundation.

\newcommand{\noopsort}[1]{}

\end{document}